\documentclass[%
prl,reprint,longbibliography,twocolumn,
 amsmath,amssymb,
 aps,
]{revtex4}

\usepackage{bm}
\usepackage{graphicx}
\usepackage{enumitem}   
\usepackage{cleveref}
\usepackage{float}

\begin{document}

\widetext

\title{First hitting times to intermittent targets}

\author{Gabriel Mercado-V\'asquez}
\author{Denis Boyer}%
 \email{boyer@fisica.unam.mx}
\affiliation{%
 Instituto de F\'isica, Universidad Nacional Aut\'onoma de M\'exico, Mexico City 04510, Mexico
}%
\date{\today}

\begin{abstract} 
In noisy environments such as the cell, many processes involve target sites that are often hidden or inactive, and thus not always available for reaction with diffusing entities. To understand reaction kinetics in these situations, we study the first hitting time statistics of a Brownian particle searching for a target site that switches stochastically between visible and hidden phases. At high crypticity, an unexpected rate limited power-law regime emerges for the first hitting time density, which markedly differs from the classic $t^{-3/2}$ scaling for steady targets. Our problem admits an asymptotic mapping onto a mixed, or Robin, boundary condition. Similar results are obtained with non-Markov targets and particles diffusing anomalously.
\end{abstract}

\maketitle

Numerous phenomena are controlled by the time taken by a process to first reach a specified target state or conformation \cite{redner2001guide,siegert1951first,gardiner2004handbook}. First passage processes allow us to understand diffusion controlled reactions \cite{szabo1980first,cui2006mean}, to predict the sizes of neuronal avalanches in neurocortical circuits \cite{beggs2003neuronal} or the search strategies adopted by foraging animals \cite{viswanathan2011physics,kagan2015search}. They are also supposed to govern the kinetics of many essential biological processes like antibody production or cell differentiation, which depend on how long it takes for fundamental steps to be completed, such as the first random encounter of two remote DNA segments \cite{zhang2016first,chou2014first}. 

The theory of first passage processes has witnessed many recent developments \cite{metzler2014first}, which usually consider fixed and perfectly absorbing targets.  
Naturally, in many cases, due to errors or imperfections in the binding or detection phase, reaction may occur only after several attempts. For instance, partial absorption is conveniently modeled through an absorption probability each time a random walker crosses a target region \cite{majumdar1998persistence,burkhardt2000dynamics,bray2013persistence}. In the Brownian limit this rule becomes equivalent to a radiation, or Robin, boundary condition, where the diffusive absorption flux is set proportional to the probability density at the target \cite{redner2001guide,singer2008partially,pal2018motion}. Nevertheless, little is known on first passage problems with barriers or targets that follow some internal dynamics\cite{benichou1999resonant,benichou2000kinetics,rojo2011intermittent}. A simple example, of interest here, is provided by switching processes between active and inactive states. 

Stochastic switching processes are inherent in cell biology. Gene expression is only possible if a binding target site along the DNA chain is accessible to diffusive transcription factors \cite{mcadams1997stochastic,tian2006stochastic}. Instead of being continuous and smooth, transcription usually occurs in bursts separated by periods of inactivity during which no transcription is carried out \cite{chubb2010bursts,suter2011mammalian}. Such transcriptional bursting is related to the chromatin remodeling state: when the chromatin is unfolded, the binding site is accessible for gene expression, whereas folded states do not allow transcription \cite{munsky2012,wu1997chromatin,eberharter2002histone}. The accessibility to the binding sites can be described by a Poisson distributed switching process, with fixed transitions rates between two states \cite{munsky2012}. In prokaryote cells such as \textit{E. coli}, the time intervals between bursty (\lq\lq on") and silent (\lq\lq off") phases are exponentially distributed. These processes can have an average duration of minutes, the periods of inactivity being much longer than active periods \cite{golding2005real}. 

Switching between bistable states also characterizes DNA looping \cite{wong2008interconvertible} (the ability of distant sites on the chain to physically interact to regulate gene expression), where the time spent in the off-state can be also very long \cite{chen2014modulation}. Similarly, ion channels stochastically transit between open and closed conformations, thus affecting transport through membranes, cell signaling or drug delivery \cite{bressloff2014stochastic,reingruber2009gated}. Studies on single ion-channels have shown that opening or closing events of the pore occur over characteristic dwell times ranging from 0.5 to hundreds of ms \cite{kawano2013automated,sakmann2013single}. These times are comparable or much larger than the diffusion times of K$^+$ or Ca$^{2+}$ ions at the scale of a cell ($\tau_D\sim 0.1\ \mu$s) \cite{donahue1988}. Hence, first hitting times are likely to be limited by the channel state \cite{lide2004crc,mashl2001hierarchical}.

\begin{figure}
\centering
			\includegraphics[width=.45\textwidth]{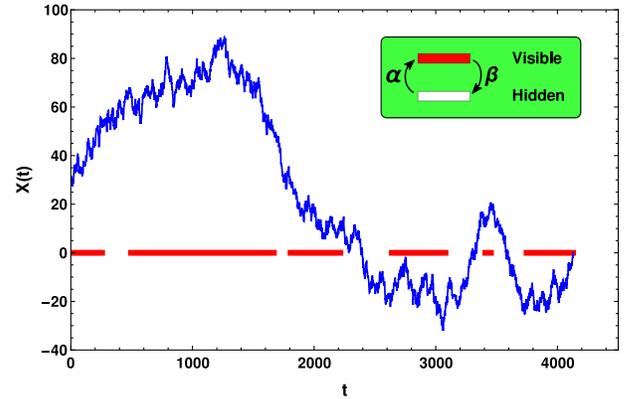}
            
			\caption{Brownian trajectory in the presence of an intermittent target located at the origin. The red segments represent the target in the visible state, separated by time intervals in the hidden state. The Brownian particle is absorbed when it reaches the target in the visible state for the first time.}
			\label{fig:walk1D}
\end{figure}

In this Letter, we address the generic but largely unexplored question of a one-dimensional unbounded Brownian motion with diffusion coefficient $D$ and an intermittent target located at the origin, as illustrated in Figure \ref{fig:walk1D}. The target internal state is characterized by a time dependent binary variable $\sigma(t)$: when $\sigma=1$ the target is visible, meaning that the Brownian particle is absorbed upon encounter; when $\sigma=0$ the target is invisible or transparent, and the particle is not absorbed when crossing the origin. The target visibility randomly switches between these two states, which last for time intervals that are exponentially distributed. The target in state $\sigma=0$ changes to state $\sigma=1$ at rate $\alpha$, whereas the reverse transition occurs at rate $\beta$ (Fig. \ref{fig:walk1D}). Therefore, the mean duration of the visible (invisible) phase is $1/\beta$ ($1/\alpha$) and the probability to find the target in the visible state is $\alpha/(\alpha+\beta)$.

With one target, this problem bears similarities with an intermittent search (IS) strategy \cite{benichou2005optimal,benichou2011intermittent}, where the searcher becomes temporarily \lq\lq blind", the target being always visible. But unlike in IS, our particle does not adopt a different transport mode in the blind phase, it just keeps diffusing. For many targets with independent internal dynamics, the two problems differ even more.

A quantity of central interest here is $Q_{\sigma_0}(x,t)$, the probability that the particle has survived up to time $t$ given that its initial position was $x$, the initial target state being $\sigma(t=0)=\sigma_0$. We set $x>0$ in the following. Averaging over the initial target states defines the average survival probability, denoted as $Q_{\rm av}(x,t)$:
\begin{equation}\label{qav}
Q_{\rm av}(x,t)=\frac{\beta}{\alpha+\beta}Q_0(x,t)+\frac{\alpha}{\alpha+\beta}Q_1(x,t).
\end{equation}
The first hitting time distribution (FHTD) $P_i$ is obtained from the usual identity $Q_i(x,t)=\int_t^{\infty}d\tau P_i(x,\tau)$ or 
\begin{equation}\label{relpq}
P_i(x,t)=-\frac{\partial Q_i(x,t)}{\partial t},
\end{equation}
where $i\in \{0,1,{\rm av}\}$. Following a method applied to nonintermittent targets and Markov processes such as intermittent search \cite{benichou2011intermittent}, run-an-tumble motion \cite{malakar2018steady} or diffusion with resetting \cite{evans2018run}, we show in the Supplemental Material (SM) that the survival probabilities satisfy two coupled backward Fokker-Planck equations:
\begin{eqnarray}
	\frac{\partial Q_0(x,t)}{\partial t}=& D\frac{\partial^2 Q_0(x,t)}{\partial x^2 }+\alpha\left[ Q_1(x,t)- Q_0(x,t)\right]\label{survivep0}\\
\frac{\partial Q_1(x,t)}{\partial t}=&D\frac{\partial^2 Q_1(x,t)}{\partial x^2 }+\beta\left[ Q_0(x,t)-Q_1(x,t)\right].
\label{survivep1}
\end{eqnarray}
These functions need to satisfy the boundary conditions
\begin{eqnarray}
Q_1(x=0,t)&=&0\label{conditionQ}\\
\frac{\partial Q_0(x,t)}{\partial x}\Big|_{x=0}&=&0.\label{conditionQzero}
\end{eqnarray}
Whereas Eq. (\ref{conditionQ}) simply asserts that the target is absorbing in the visible state, relation (\ref{conditionQzero}) is a bit more subtle. It can be understood, for instance, by considering the case $\beta=0$ and a target in state $0$ at $t=0$, which therefore irreversibly transits to the visible state at rate $\alpha$. The calculation of $Q_0$ in this case is performed in the SM by using simple probabilistic arguments. One checks {\it a posteriori} that the solution fulfills condition (\ref{conditionQzero}). Similar arguments allow us to show that Eq. (\ref{conditionQzero}) holds in the general case $\beta>0$ as well (see the SM).

Let us define the Laplace transforms   $\widetilde{Q}_{\sigma_0}(x,s)=\int^{\infty}_0Q_{\sigma_0}(x,t)e^{-st}dt$, which satisfy the following system
\begin{eqnarray}
D\frac{\partial^2 \widetilde{Q}_0(x,s)}{\partial x^2}+\alpha\widetilde{Q}_1(x,s)-(\alpha+s)\widetilde{Q}_0(x,s)=-1
\label{eqtilde0}\\
D\frac{\partial^2 \widetilde{Q}_1(x,s)}{\partial x^2}+\beta\widetilde{Q}_0(x,s)-(\beta+s)\widetilde{Q}_1(x,s)=-1
\label{eqtilde1}
\end{eqnarray}
The general solutions are $\widetilde{Q}_0(x,s)=Ae^{\pm a\sqrt{s}}-\alpha Be^{\pm a\sqrt{s+\alpha+\beta}}/\beta+1/s$ and $\widetilde{Q}_1(x,s)=Ae^{\pm a\sqrt{s}}+Be^{\pm a\sqrt{s+\alpha+\beta}}+1/s$, where we have employed the notation
\begin{equation}\label{defa}
a=x/\sqrt{D}. 
\end{equation}
Thus, $\tau_D\equiv a^2$ is the typical diffusion time to reach the target region. With the boundary conditions (\ref{conditionQ})-(\ref{conditionQzero}) and noting that $Q_{\sigma_0}$ must remain finite as $a\rightarrow \infty$, one deduces 
\begin{eqnarray}
\widetilde{Q}_i(x,s)=&&-\frac{\alpha\sqrt{s+\alpha+\beta}}{\sqrt{s}\left(\alpha\sqrt{s+\alpha+\beta}+\beta\sqrt{s}\right)}\Bigg(\frac{e^{-a\sqrt{s}}}{\sqrt{s}}\nonumber\\
&&-C_i\frac{e^{-a\sqrt{s+\alpha+\beta}}}{\sqrt{s+\alpha+\beta}}\Bigg)+\frac{1}{s}
\label{generalQ}
\end{eqnarray}
where, again, $i=\{0,1,{\rm av}\}$ and the constants $C_i$ take the  values $C_0=1$, $C_1=-\frac{\beta}{\alpha}$ and $C_{av}=0$.

The Laplace transform of the FHTD is deduced from the general relation $\widetilde{P}_i(x,s)=1-s\widetilde{Q}_i(x,s)$, which stems from Eq. (\ref{relpq}). 
The solutions for $\widetilde{Q}_i$ or $\widetilde{P}_i$ do not seem to have simple inverses, nevertheless Eq. (\ref{generalQ}) can be inverted by means of the convolution theorem and the complete solution expressed in a rather lengthy integral form. This exact solution is given in the SM.   

With the Laplace expressions (\ref{generalQ}) at hand, one easily checks that in the limit $\beta\rightarrow0$, one recovers the well-known case of a target always in the visible state:
\begin{equation}
\widetilde{Q}_{{\rm av}}(x,s)\rightarrow
\widetilde{Q}^{{\rm st}}(x,s)=\frac{1-e^{-a\sqrt{s}}}{s},\label{qst}
\end{equation}
or $\widetilde{P}^{\rm st}(x,s)=e^{-a\sqrt{s}}$, where the label \lq\lq st" stands for the standard case of a nonintermittent target \cite{redner2001guide}. This expression is inverted as the L\'evy-Smirnov distribution 
\begin{equation}\label{pst}
P^{st}(a,t)=\frac{a}{\sqrt{4\pi t^3}}e^{-\frac{a^2}{4t}}\simeq \frac{a}{\sqrt{4\pi}}\ t^{-\frac{3}{2}}\ \ 
{\rm for}\ t\gg \tau_D.
\end{equation}

For very large values of the two transition rates compared to the inverse diffusion time $1/a^2$, and keeping $\beta/\alpha$ constant, Eq. (\ref{generalQ}) shows that the three FHTDs also tend to that of the standard problem:
\begin{equation}
{Q}_{i}(x,t)\rightarrow{Q}^{st}(x,t).
\end{equation} 
This result may seem counterintuitive, as it tells that at very high transition rates, the target is easily detectable by the Brownian particle, \emph{even if it is invisible most of the time}, {\it i.e.}, with $\beta/\alpha$ fixed to a large value.
The fast absorption can be understood here by the recurrence of Brownian trajectories in $1d$: a particle crossing the origin re-crosses it many times within a short period. If, in the meantime, the target rapidly transits from one state to the other, as soon as it becomes visible, it will be hit by the nearby particle. A similar mechanism can explain the fast absorption of ligands that spend most of the time in a hidden state \cite{reingruber2009gated}. 

We next comment on a key property which is not met with steady targets or in usual radiation boundary problems. In the limit of infinitely fast diffusion, $D\rightarrow\infty$ or $a\rightarrow0$, a steady target is found immediately [$Q^{st}(a=0,t)=0$], since Brownian motion is recurrent in $1d$. In contrast, $Q_{\rm av}$ and $Q_0$ admit non-trivial limits for intermittent targets. Defining $Q^{I}_{i}(t)\equiv Q_{i}(a=0,t)$, the survival probability for any $a$ can be decomposed as:
\begin{equation}\label{decomp}
    Q_{i}(a,t)=Q^I_{i}(t)+Q^{D}_{i}(a,t)
\end{equation}
where $i=\{0,{\rm av}\}$.
By construction, $Q^{D}_{i}(a,t)$ vanishes as $a\rightarrow0$ and represents the diffusion limited part of the survival probability, whereas $Q^{I}_{i}(t)$, which depends only on $\alpha$ and $\beta$, is the contribution limited by the target dynamics. This intermittent part arises from the fact that the target can be initially invisible and therefore undetectable while it remains so, no matter how fast diffusion occurs. $Q^I_{0}(t)$ is the probability that the particle starting right at the position of the initially invisible target, has still not hit it at $t$. Owing to Eq. (\ref{relpq}), $P_{i}(a,t)$ can also be decomposed as $P^I_{i}(t)+P^{D}_{i}(a,t)$. From Eq. (\ref{generalQ}), one obtains in the Laplace domain
\begin{equation}
\widetilde{Q}^I_{0}(s)=\frac{\beta+\alpha}{\sqrt{s}(\alpha\sqrt{s+\alpha+\beta}+\beta\sqrt{s})},\label{QI}
\end{equation}
whereas $\widetilde{Q}_{\rm av}^I(s)=\frac{\beta}{\alpha+\beta}\widetilde{Q}^I_{0}(s)$. 

\begin{figure}
\centering \includegraphics[width=0.48\textwidth]{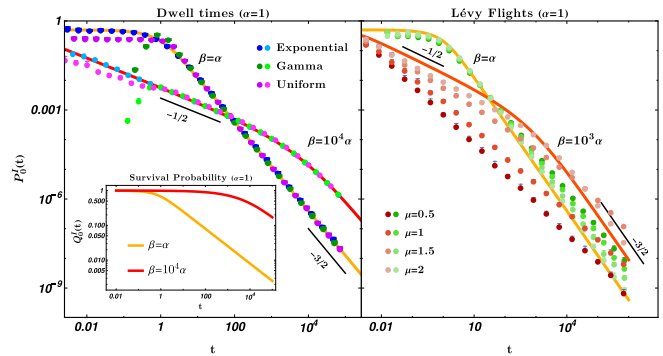}
\caption{First hitting time density for a particle starting at the origin. Symbols represent simulation results and lines the exact solution obtained from Eq. (\ref{QI}). {\bf Left:} Brownian particle and several distributions of activity/inactivity times:
exponential, Gamma (with shape parameter $5$) and uniform. In each case, the mean time of the \lq\lq off" (\lq\lq on") state is set to $1/\alpha$  ($1/\beta$, respectively). {\bf Right:} Discrete L\'evy flights of index $\mu$ and time-step $10^{-4}$, with an exponential target.}
\label{fig:LogLogSimu}
\end{figure}

We now show how this intermittent contribution drastically affects the asymptotic properties of the FHTD, especially in the {\it cryptic} regime $\beta/\alpha\gg 1$. Since $\widetilde{Q}_{av}(a,s=0)=\infty$, the mean first hitting time is infinite like in the standard case, and we can deduce the large time behaviors from a small $s$ expansion. Approximating $\sqrt{s+\alpha+\beta}$ by $\sqrt{\alpha+\beta}$, Eq. (\ref{generalQ}) is recast as 
\begin{equation}
    \widetilde{Q}_{\rm av}(a,s)\simeq\frac{1}{s}\left(1-\frac{e^{-a\sqrt{s}}}{1+K\sqrt{s}}\right)
    \label{Qaveapprox}
\end{equation}
with $K=\beta/(\alpha\sqrt{\alpha+\beta})$. The right-hand-side of Eq. (\ref{Qaveapprox}) can be exactly inverted \cite{abramowitz1965handbook} and combined with Eq. (\ref{relpq}), yielding:
\begin{equation}
\begin{aligned}
    P_{\rm av}(a,t)\simeq&\frac{1}{K\sqrt{\pi t}}\exp\left(-\frac{a^2}{4t}\right)\\
    &-\frac{1}{K^2}\operatorname{erfc}\left(\frac{\sqrt{t}}{K}+\frac{a}{2\sqrt{t}}\right)\exp\left(\frac{a}{K}+\frac{t}{K^2}\right).
    \label{PavLargeT}
    \end{aligned}
\end{equation}
Eq. (\ref{PavLargeT}) is valid for times larger than the target time-scale defined as $\tau_{ta}\equiv(\alpha+\beta)^{-1}$. It is very close to the exact solution obtained by convolution in the SM for all $t$ (see Fig. \ref{fig:LogLogPav3}). Two scaling regimes emerge, as can be seen directly from Eq. (\ref{Qaveapprox}) by setting $a\sqrt{s}\ll1$: 

\begin{enumerate}[label=\bfseries(\roman*)]
\item For $K\sqrt{s}\ll 1$, one has $\widetilde{Q}_{\rm av}\simeq (a+K)/\sqrt{s}$, which is inverted as $Q_{\rm av}(a,t)\simeq (a+K)/\sqrt{\pi t}$. Hence, with Eq. (\ref{relpq}), the asymptotic scaling (\ref{pst}) generalizes to 
\begin{equation}\label{scal32}
P_{\rm av}(a,t)\simeq (a+K)/\sqrt{4\pi t^3}.
\end{equation}

\begin{figure}
\centering \includegraphics[width=0.48\textwidth]{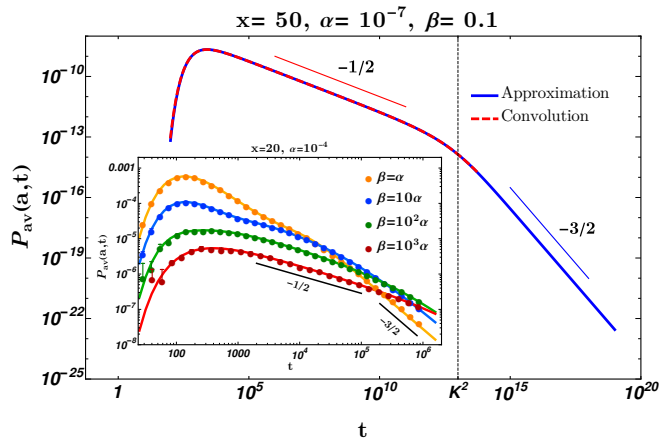}
\caption{Average first hitting time density in a cryptic case. The searcher starting position is $x=50$, and $D=1/2$. The crossover time $K^2$ is $\approx10^{13}$ in this example. Inset: Average FHTD for $\alpha=10^{-4}$ and varying $\beta$. Symbols represent simulation results and lines the exact solution.}
\label{fig:LogLogPav3}
\end{figure}

\item If $K\gg a$, an intermediate regime is possible, where $a\sqrt{s}\ll1$ but $K\sqrt{s}\gg 1$ . In this case $\widetilde{Q}_{\rm av}\simeq\frac{1}{s}\left[1-1/(K\sqrt{s})\right]$, or  $\widetilde{P}_{\rm av}=1-s\widetilde{Q}_{\rm av}\simeq 1/(K\sqrt{s})$. Hence,  
\begin{equation}\label{scal12}
P_{\rm av}(a,t)\simeq 1/(K\sqrt{\pi t}).  
\end{equation}
\end{enumerate}
Eq. (\ref{scal12}) is one of our main results.
From the above considerations, $\tau_c\equiv K^2$ sets a crossover time separating the standard $t^{-3/2}$ scaling (with a modified prefactor) from a new intermediate regime with exponent $-1/2$, holding in the range ${\rm max}(\tau_{ta},\tau_D)\ll t\ll \tau_c$.
This regime is intermittency dominated, as Eq. (\ref{scal12}) does not involve $a$. Clearly, it can be observed only if $\tau_c/\tau_{ta}\gg 1$. As
\begin{equation}
    \tau_c/\tau_{ta}=(\beta/\alpha)^2,
\end{equation}
the intermediate region exists for $\beta/\alpha\gg1$, at high crypticity, and
its extent rapidly increases with $\beta/\alpha$. As shown by Figure \ref{fig:LogLogPav3}, this scaling law can span many decades, broadening considerably the FHTD and making the standard regime hard to reach. Meanwhile, $Q_{\rm av}(a,t)\simeq 1-(2/\sqrt{\pi})\sqrt{t/\tau_c}$ remains close to unity [see inset of Fig. \ref{fig:LogLogSimu}-Left], {\it i.e.}, target encounters are very rare. 

At short times ($t\ll\tau_{ta}$), Eq. (\ref{PavLargeT}) is not valid and the FHTD can be deduced from a large $s$ expansion. Setting $a=0$ for simplicity, one gets from Eq. (\ref{QI}):
$\widetilde{Q}^I_{\rm av}(s)\simeq \frac{\alpha}{(\alpha+\beta)s}\left[\frac{\beta}{\alpha }-\frac{\beta}{2s}+\frac{\beta(3\alpha+\beta)}{8s^2}+\cdots\right]$, which by inversion yields
\begin{equation}
    P^I_{\rm av}(t)\simeq\frac{\alpha}{\alpha+\beta}\left[\delta(t)+\frac{\beta}{2}-\frac{\beta(3\alpha+\beta)}{8}t+\cdots\right].
\end{equation}

The exact solution obtained in the SM is checked successfully with Monte Carlo simulations in Figure \ref{fig:LogLogPav3}-Inset for several crypticity strengths. The intermediate regime is already noticeable at $\beta/\alpha\approx 10$. 
Note that biological systems are often cryptic and with $\tau_c\gg \tau_D$. Transcriptional bursting in prokaryotic cells is characterized by relatively short periods during which transcription is allowed, corresponding to $\beta/\alpha\approx 6$ \cite{golding2005real}. Lactose repressors can also form long-tether DNA loops (that block transcription) at a rate $10$ times faster in than the loop$\rightarrow$unlooped transition \cite{wong2008interconvertible}. The activity times $\tau_c$ in these examples are of the order of minutes, much larger than $\tau_D\sim 1$ s for a protein in a cell \cite{goulian2000tracking}. In presynaptic processes, the parametrized Hodgkin--Huxley model predicts a $\beta/\alpha$ of $\approx  4500$ for the switching rate to the inactivated state of Na$^+$ channels at  rest voltage ($50$ \textit{mV})\cite{bressloff2014stochastic,ermentrout2010mathematical}. 

Since the target switches between absorbing and hidden phases (the latter being equivalent to reflecting in the present geometry), one may wonder about a possible connection with diffusion in the presence of a mixed, or Robin, boundary condition (RBC). Let $p(z,t)$ be the probability density of the position 
$z\in[0,\infty)$ of a Brownian particle with a RBC at the origin, namely \cite{singer2008partially}:
\begin{equation}\label{Robin}
    \left. D\frac{\partial p}{\partial z}\right|_{z=0}=\kappa p(z=0,t),
\end{equation}
where $\kappa$ is a positive constant. Eq. (\ref{Robin}) is widely used in effective medium descriptions of spatially heterogeneous interfaces containing both reflecting and reactive zones \cite{zwanzig1990diffusion,batsilas2003stochastic,berezhkovskii2004boundary}. The exact survival probability in $1d$ of a particle starting at $z=x$ and obeying a RBC actually coincides with our Eq. (\ref{Qaveapprox}) or (\ref{PavLargeT}) for all $t$, where $K$ must be replaced by $\sqrt{D}/\kappa$ \cite{sano1979partially,pal2018motion}. Consequently, as far as survival is concerned, both problems become equivalent at large times. We deduce the formula
\begin{equation}\label{kappa}
\kappa=\frac{\alpha}{\beta}\sqrt{\alpha+\beta}\sqrt{D}.
\end{equation}
As one may expect, the boundary is absorbing ($\kappa\rightarrow\infty$) when $\beta\rightarrow 0$ and reflecting ($\kappa\rightarrow0$) when $\alpha\rightarrow 0$. Non trivially, it is also absorbing as $\alpha,\beta\rightarrow\infty$, $\beta/\alpha$ being fixed, as mentioned earlier. The two-state process thus provides a new, rigorous example of application of the RBC (\ref{Robin}), extending the relevance of the latter to the study of fluctuating biophysical systems. Both problems differ for $t$ smaller than the target time-scale, though, as the RBC does not involve such a time-scale. A similar asymptotic analogy with the RBC was shown some time ago for diffusion into a partially absorbing medium \cite{ben1993partial}. 

We discuss the generality of our findings when more complex processes come into play.  On-off processes such as DNA looping or ion channel dynamics can be substantially non-Markov \cite{colquhoun1981fluctuations,liebovitch1987ion,goychuk2004fractional}. We have simulated targets with activity and inactivity times that were distributed non-exponentially in several ways. As shown by Figure \ref{fig:LogLogSimu}-Left, the scaling regimes (\ref{scal32}) and (\ref{scal12}) still hold in those cases, and our previous solution remains quantitatively correct at large times [$t>$max$(1/\alpha,1/\beta$)]. 

The Brownian motion case can also be extended to anomalous transport. Given a Markovian target, we simulated searchers performing one step per time unit ($\Delta t\ll 1/\alpha$ and $1/\beta$), {\it i.e.}, with position $X_n=\sum_{i=1}^n\ell_i$ with $t=n\Delta t$, and where the i.i.d. $\ell_i$'s follow a symmetric L\'evy stable distribution of index $0<\mu<2$ \cite{chambers1976method}. The process terminates when $X_n$ changes sign while the target is active. In the border case $\mu=2$, the distribution of the $\ell_i$'s was $\propto|\ell|^{-3}$. For $\beta\sim\alpha$, the FHTD for a particle starting at the origin (with $\sigma_0=0$) depends surprisingly little on $\mu$ and is close to the Brownian curve, see  Fig. \ref{fig:LogLogSimu}-Right. Although approximate, this independence is reminiscent of the universality of the Sparre Andersen theorem, valid for any unbiased continuous $1d$ process being absorbed when first crossing the origin \cite{feller2008introduction}. If $\beta\gg\alpha$, the process crosses the origin many times before absorption. The intermediate regime appears, as for Brownian motion. However, the corresponding exponent $\zeta_\mu$ continuously depends on $\mu$: $\zeta_2\approx -1/2$ whereas $\zeta_{1}\approx\zeta_{1/2}\approx -1$. Except for $\mu=2$, the $-3/2$ regime was not observed as it may settle at very large times.

In summary, we have shown that diffusive search processes can be severely affected by the intermittent switching dynamics of a target site, a situation often met in noisy complex media. A new, rate controlled scaling regime with exponent $-1/2$ emerges at high target crypticity, and the problem can be mapped onto a radiation boundary problem at large times.  These results can be readily extended to higher spatial dimensions with the same formalism. Our findings point toward intermittent dynamics as a way of regulating first passage processes in the cell. They can also have implications in foraging ecology, where animals are able to be cryptic and undetectable by predators for long period of times by camouflaging themselves \cite{stevens2009animal}, or adopting a subterranean lifestyle 
\cite{edmunds1990evolution,ruxton2004avoiding,gendron1983searching,o1990search}. According to Eq. (\ref{kappa}), to avoid predators and fulfill the constraint of spending a certain fraction of time outside, animals should space out consecutive exits in time, a behaviour actually observed in female ground squirrels \cite{williams2014light,williams2016secret}.
Macroscopic search experiments with dynamical targets can be achieved by means of mobile robots with a limited sensing range and fixed sources emitting intermittent electromagnetic signals \cite{song2011stochastic,song2011time}. We finally mention that the decay of the survival probability in unconfined space plays an important role for the large volume scaling of the FHTD in confined environments \cite{levernier2018universal}. The intermediate regime for cryptic targets should thus have important consequences for confined walks \cite{condamin2007first}.

We thank Germinal Cocho and PAPIIT-DGAPA through Grant IN108318 for support. G. M. V. thanks CONACYT for scholarship support. We thank O. B\'enichou, P. Cluzel, A. Kundu, S. N. Majumdar, P. Miramontes, I. P\'erez-Castillo, G. Schehr and F. J. Sevilla for fruitful discussions.


%

\newpage

\widetext
\begin{center}
{\large Supplemental Material: First hitting times to intermittent targets}

\vspace{0.5cm}
Gabriel Mercado-V\'asquez and Denis Boyer
\end{center}

\vspace{0.5cm}
\section{Governing equations}

We derive here the coupled backward Fokker-Planck equations satisfied by the survival probabilities $Q_{0,1}(x,t)$ and which read:
\begin{eqnarray}
	\frac{\partial Q_0(x,t)}{\partial t}=D\frac{\partial^2 Q_0(x,t)}{\partial x^2 }+\alpha[Q_1(x,t)-Q_0(x,t)]\label{survivep0}\\
\frac{\partial Q_1(x,t)}{\partial t}=D\frac{\partial^2 Q_1(x,t)}{\partial x^2 }+\beta[Q_0(x,t)-Q_1(x,t)]
\label{survivep1}
\end{eqnarray} 

We suppose that the Brownian particle starts from a position $x>0$ at time $t=0$, with the target in the initial state $\sigma_0=0$. We notice that during a small time interval $[0,\Delta t]$, the target can switch to the state $\sigma=1$ with probability $\alpha \Delta t$ or remain in $\sigma=0$ with probability $1-\alpha\Delta t$. Summing these two contributions, one obtains a relation for the quantity $Q_0(x,t+\Delta t)$:

\begin{equation}
 	Q_0(x,t+\Delta t)=\alpha\Delta t\int_{-\infty}^{\infty}d\xi P_{\Delta  t}(\xi)Q_1(x+\xi,t)
 	+(1-\alpha\Delta t)\int_{-\infty}^{\infty}d\xi P_{\Delta t}(\xi)Q_0(x+\xi,t)
\end{equation}
At  time $\Delta t$, the particle position is $x+\xi$, which serves as a new initial condition for the rest of the trajectory in the interval $[\Delta t,t+\Delta t]$, which is of duration $t$. The random displacement $\xi$ of the Brownian particle is drawn from the Gaussian distribution $P_{\Delta t}(\xi)$ with zero mean and variance $\langle \xi^2 \rangle=2D\Delta t$. A Taylor expansion of the probabilities $Q_{0,1}(x+\xi,t)$ in powers of $\xi$ up to second order (the first non-zero average contribution) gives:
\begin{equation}
    Q_{0,1}(x+\xi,t)=Q_{0,1}(x,t)+\xi 
    \frac{\partial Q_{0,1}(x,t)}{\partial x}
    +\frac{\xi^2}{2}
    \frac{\partial^2 Q_{0,1}(x,t)}{\partial x^2}
    +...
\end{equation}
Similarly $Q_0(x,t+\Delta t)= Q_0(x,t)+\Delta t \frac{\partial Q_0(x,t)}{\partial t}+...$
After using the fact that $\langle \xi\rangle=0$ and $\langle \xi^2 \rangle=2D\Delta t$, we group the terms of order $\Delta t$, take the limit $\Delta t\rightarrow0$, and Eq. (\ref{survivep0}) is obtained. Eq. (\ref{survivep1}) is deduced from a similar reasoning, noting that now the probability to switch from the state $\sigma=1$ to $\sigma=0$ during the time interval $[0,\Delta t]$ is $\beta\Delta t$, whereas, with probability $1-\beta\Delta t$, the target will remain with $\sigma=1$.

\section{Boundary condition for $Q_0(x,t)$ 
}

We suppose that at $t=0$ the target site, which fixed at $x=0$, is in the dormant state $\sigma_0=0$. The probability that the target switches to the active state $\sigma=1$ for the first time at time $t'$ is given by the exponential distribution with rate $\alpha$. If we choose $\beta=0$, once the target has switched to the active state, it will remain active for ever. Since the Brownian particle started at $x>0$ at $t=0$ and diffuses in the unbounded free space, it will be located at a Gaussianly distributed position $x'$ at time $t'$. This position $x'$ represents a new initial position for the standard first passage problem with a permanent target. For $\beta=0$, the survival probability at time $t$ will thus be given by
\begin{equation}
    Q_0(x,t)=\int_0^\infty dt' \alpha e^{-\alpha t'}\int_{-\infty}^\infty dx' \frac{e^{-\frac{(x'-x)^2}{4Dt'}}}{\sqrt{4\pi Dt'}}
    \times\begin{cases}
    1 &\text{if  $t'>t$}\\
    Q^{st}(x',t-t') &\text{if  $t'<t$}
    \end{cases}    
        \label{qbetazero}
\end{equation}
where $Q^{st}(x,t)$ stands for the standard survival probability with a steady target. In Eq. (\ref{qbetazero}), we have used the fact that the particle cannot be absorbed while the target is inactive ($t<t'$), and the averages over $x'$ and $t'$ are taken. Let us also take the Laplace transform of Eq. (\ref{qbetazero}), denoting $\widetilde{Q}_0(x,s)=\int_0^\infty dt\  e^{-st}Q_0(x,t)$. After changing the order of the integrals, we obtain
\begin{equation}
    \widetilde{Q}_0(x,s)=\alpha\int_0^\infty dt' e^{-\alpha t'}\int_{-\infty}^\infty dx' \frac{e^{-\frac{(x'-x)^2}{4Dt'}}}{\sqrt{4\pi Dt'}}
    \left( \int_0^{t'}dt\  e^{-st}+\int_{t'}^\infty dt\  e^{-st}Q^{st}(x',t-t')\right)
\label{qtilde2}
\end{equation}
The first integral of the second line of Eq. (\ref{qtilde2}) is simply $\frac{1-e^{-st'}}{s}$ and does not depend on $x'$, thus, the integral over $x'$ gives unity. For the second integral of the same line, we make a change of variable $u=t-t'$ and obtain the Laplace transform of $Q^{st}(x,u)$ multiplied by a factor $e^{-st'}$. Therefore:
\begin{equation}
\begin{aligned}
   \widetilde{Q}_0(x,s) =&\alpha\int_0^\infty dt' e^{-\alpha t'}\Bigg(\frac{1-e^{-st'}}{s}
   +e^{-st'}\int_{-\infty}^\infty dx' \frac{e^{-\frac{(x'-x)^2}{4Dt'}}}{\sqrt{4\pi Dt'}}\widetilde{Q}^{st}(x',s)\Bigg)\\
   =&\frac{1}{s+\alpha}+\alpha\int_{-\infty}^\infty dx' \frac{e^{-|x'-x|\sqrt{(s+\alpha)/D}}}{\sqrt{4D(s+\alpha)}}\widetilde{Q}^{st}(x',s)
    \end{aligned}\label{qtilde3}
\end{equation}
Given that $\widetilde{Q}^{st}(x',s)=\frac{1-e^{-|x'|\sqrt{s/D}}}{s}$ 
\cite{redner2001guide} we substitute this expression into Eq. (\ref{qtilde3}) and obtain 
\begin{equation}
   \widetilde{Q}_0(x,s)=\frac{1}{s+\alpha}+\frac{\alpha}{2s{\sqrt{D(s+\alpha)}}}
   \int_{-\infty}^\infty dx' e^{-|x'-x|\sqrt{(s+\alpha)/D}}\left(1-e^{-|x'|\sqrt{s/D}}\right)
    \label{qtilde4}
\end{equation}
After integrating in Eq. (\ref{qtilde4}), we obtain
\begin{equation}\label{q0final}
    \widetilde{Q}_0(x,s)=\frac{1}{s}+\frac{\sqrt{s}e^{-x\sqrt{(s+\alpha)/D}}-\sqrt{s+\alpha}e^{-x\sqrt{s/D}}}{s\sqrt{s+\alpha}}.
\end{equation}
This result coincides with Eq. (10) in the main text for the case $i=0$ and $\beta=0$. It is also easy to see that Eq. (\ref{q0final}) satisfies the boundary condition
\begin{equation}
\frac{\partial \widetilde{Q}_0(x,s)}{\partial x}\Big|_{x=0}=0,\label{conditionQzero}
\end{equation}
for any $s$, which implies that
\begin{equation}
\frac{\partial {Q}_0(x,t)}{\partial x}\Big|_{x=0}=0,
\end{equation}
for any $t$.
Hence, it is \lq\lq as if" the boundary in $x=0$ was always reflective, given its initial reflecting state $\sigma=0$. This may look surprizing, since the target is actually reactive (absorbing) any time after the switch time $t'$. 

In the general case $\beta\neq0$, the boundary condition can actually be derived in a similar way. As before, we suppose that the initial target state is $\sigma_0=0$. The particle starts from the position $x>0$ and freely diffuses until the first transition to the state $\sigma=1$ occurs, at a time $t'$. During the interval $[0,t']$ the particle has reached the Gaussianly distributed position $x'$. At this point, the deduction departs from the case of $\beta=0$; now, the survival probability function for times $t>t'$ is no longer the standard survival probability but $Q_1(x',t-t')$, since a renewed diffusion process starts from $x'$ with $\sigma_0=1$ at time $t'$. Thus, the survival probability at time $t$ will be
\begin{equation}
    Q_0(x,t)=\int_0^\infty dt' \alpha e^{-\alpha t'}\int_{-\infty}^\infty dx' \frac{e^{-\frac{(x'-x)^2}{4Dt'}}}{\sqrt{4\pi Dt'}}
    \times\begin{cases}
    1 &\text{if  $t'>t$}\\
    Q_1(x',t-t') &\text{if  $t'<t$}
    \end{cases}
        \label{qanybeta}
\end{equation}
In Eq. (\ref{qanybeta}), $Q_1$ does not depend on the initial position $x$, thus the derivative with respect to $x$ will only apply to the exponential term in the integral. Evaluating $\partial Q_0 (x,t)/\partial x$ in $x=0$ gives
\begin{equation}
    \frac{\partial Q_0(x,t)}{\partial x}\Big|_{x=0}=\int_0^\infty dt' \alpha e^{-\alpha t'}\int_{-\infty}^\infty dx' \frac{x'e^{-\frac{x'^2}{4Dt'}}}{2Dt'\sqrt{4\pi Dt'}}
    \times\begin{cases}
    1 &\text{if  $t'>t$}\\
    Q_1(x',t-t') &\text{if  $t'<t$}
    \end{cases}
        \label{partialqanybeta}
\end{equation}
From the parity of the Gaussian function and of $Q_1$ with respect to $x'$, namely, $Q_1(-x',t-t')=Q_1(x',t-t')$, we see that the integral over $x'$ in Eq. (\ref{partialqanybeta}) vanishes. Hence, we have shown that the boundary condition of Eq. (6) in the main text is valid for any $\beta$.   


On the other hand, all the Monte Carlo simulations are in perfect agreement with the theoretical calculations, giving further support to the validity of the boundary conditions. 

\section{Solution in the Laplace domain}

In the Laplace domain, from Eqs. (\ref{survivep0})-(\ref{survivep1}) we obtain the following system:
\begin{eqnarray}
D\frac{\partial^2 \widetilde{Q}_0(x,s)}{\partial x^2}+\alpha\widetilde{Q}_1(x,s)-(\alpha+s)\widetilde{Q}_0(x,s)=-1
\label{eqtilde0}\\
D\frac{\partial^2 \widetilde{Q}_1(x,s)}{\partial x^2}+\beta\widetilde{Q}_0(x,s)-(\beta+s)\widetilde{Q}_1(x,s)=-1
\label{eqtilde1}
\end{eqnarray}
where the initial condition $Q_{\sigma}(x,t=0)=1$ has been used. The homogeneous part of Eqs. (\ref{eqtilde0})-(\ref{eqtilde1}) admits solutions of the form $q_{0}e^{\lambda x}$ and  $q_{1}e^{\lambda x}$ which satisfy
\begin{equation}
\begin{pmatrix}
	D\lambda^2-(\alpha+s) & \alpha \\
	 \beta & D\lambda^2-(\beta+s)
	\end{pmatrix} 
    \begin{pmatrix}
    q_0 \\
    q_1
    \end{pmatrix}= 0.
\end{equation}
The eigenvalues are
\begin{equation}
\lambda_1=\pm\sqrt{\frac{s}{D}}, \quad \lambda_2=\pm\sqrt{\frac{s+\alpha+\beta}{D}},
\end{equation}
with their corresponding eigenstates:
\begin{equation}
	v_1=\begin{pmatrix}
	1\\
	1
	\end{pmatrix}, \quad v_{2}=\begin{pmatrix}
	-\frac{\alpha}{\beta}\\
	1
	\end{pmatrix}.
\end{equation}
The inhomogeneous part of $\widetilde{Q}_{\sigma}$ is independent of $x$ and simply given by $1/s$ for $\sigma=0$ and $1$. Setting $x>0$, one notices that $\tilde{Q}_{\sigma}(x,s)$ must tend to $1/s$ as $x\rightarrow\infty$, since $Q_{\sigma}(x,t)$ remains equal to unity when the particle is very far from the target. Hence, only the negative eigenvalues are acceptable. One deduces
\begin{eqnarray}
\tilde{Q}_0(x,s)=Ae^{-\sqrt{\frac{s}{D}}x}-\frac{\alpha}{\beta}Be^{-\sqrt{\frac{s+\alpha+\beta}{D}}x}+\frac{1}{s}\\
\tilde{Q}_1(x,s)=Ae^{-\sqrt{\frac{s}{D}}x}+Be^{-\sqrt{\frac{s+\alpha+\beta}{D}}x}+\frac{1}{s},
\end{eqnarray}
with $A$ and $B$ two constants. Let us use the notation 
\begin{equation}\label{defa}
    a=\frac{x}{\sqrt{D}}.
\end{equation}
Enforcing the boundary condition $Q_1(x=0,t)=0$ as well as that given by Eq. (\ref{conditionQzero}) yields
\begin{eqnarray}
A=-\frac{\alpha\sqrt{s+\alpha+\beta}}{s\left(\alpha\sqrt{s+\alpha+\beta}+\beta\sqrt{s}\right)}\\
B=-\frac{\beta}{\sqrt{s}\left(\alpha\sqrt{s+\alpha+\beta}+\beta\sqrt{s}\right)}.
\end{eqnarray}
%
%
The solutions are thus given by
\begin{equation}
\widetilde{Q}_i(a,s)=-\frac{\alpha\sqrt{s+\alpha+\beta}}{\sqrt{s}\left(\alpha\sqrt{s+\alpha+\beta}+\beta\sqrt{s}\right)}\Bigg(\frac{e^{-a\sqrt{s}}}{\sqrt{s}}
-C_i\frac{e^{-a\sqrt{s+\alpha+\beta}}}{\sqrt{s+\alpha+\beta}}\Bigg)+\frac{1}{s}.
\label{generalQ}
\end{equation}
where $i=\{0,1,av\}$ and the constants $C_i$ take the  values $C_0=1$, $C_1=-\frac{\beta}{\alpha}$ and $C_{av}=0$. Hence, the average survival probability, of particular interest in this study, reads
\begin{equation}\label{qav}
\widetilde{Q}_{av}(a,s)=\frac{1}{s}\left(1
-\frac{\alpha\sqrt{s+\alpha+\beta}}{\alpha\sqrt{s+\alpha+\beta}+\beta\sqrt{s}}e^{-a\sqrt{s}}
\right),
\end{equation}
whereas $\widetilde{P}_{av}=1-s\widetilde{Q}_{av}$ is given by
\begin{equation}\label{pav}
\widetilde{P}_{av}(a,s)=
\frac{\alpha\sqrt{s+\alpha+\beta}}{\alpha\sqrt{s+\alpha+\beta}+\beta\sqrt{s}}e^{-a\sqrt{s}},
\end{equation}

\section{Exact expressions for the $Q_i(x,t)$'s as convolutions}

The general expressions (\ref{generalQ})-(\ref{pav}) do not seem to admit exact inverse Laplace transforms in terms of elementary functions. We can nevertheless obtain $Q_i(a,t)$ by using the convolution theorem \cite{spiegel1965laplace}. To this end, we write Eq. (\ref{generalQ}) in terms of a product of two functions [Eqs. (\ref{fs}-\ref{gjs}) below] and then determine its inverse through a convolution. We define:
\begin{eqnarray}
f(s)=\frac{\sqrt{s}-\frac{\beta}{\alpha}\sqrt{s+\alpha+\beta}+\frac{\alpha+\beta}{\sqrt{s}}}{(\alpha-\beta)s+\alpha^2}\label{fs}\\ 
g_i(a,s)=\frac{e^{-a\sqrt{s}}}{\sqrt{s}}-C_i\frac{e^{-a\sqrt{s+\alpha+\beta}}}{\sqrt{s+\alpha+\beta}},\label{gjs}
\end{eqnarray}
which are such that $Q_i(a,s)=1/s-\frac{\alpha^2}{\alpha+\beta}f(s)g_i(a,s)$. Denoting the inverse Laplace transforms of $f(s)$ and $g_i(a,s)$ as $F(t)$ and $G_i(a,t)$, respectively, we have: 
\begin{equation}
Q_i(a,t)=1-\frac{\alpha^2}{\alpha+\beta}\int_0^tF(u)G_i(a,t-u) du.
\label{convolutionQ}
\end{equation}
The first hitting time distribution $P_i(a,t)$, is deduced from the general identity $P_i=-\partial Q_i/\partial t$, or:
\begin{equation}
P_i(a,t)=\frac{\alpha^2}{\alpha+\beta}\int_0^tF(u)\frac{\partial G_i(a,t-u)}{\partial t} du\label{convolution}
\end{equation}
where we have use the fact that $F(t)G_i(a,0^{+})=0$.
The function $G_i(a,t)$ can be found by direct Laplace inversion \cite{abramowitz1965handbook}: 
\begin{equation}
G_i(a,t)=\frac{1}{\sqrt{\pi t}}e^{-\frac{a^2}{4t}}\left(1-C_i e^{-(\alpha+\beta) t}\right).\label{generalG}
\end{equation}
For $F(t)$, it is necessary to distinguish two cases: $\alpha=\beta$ and $\alpha\neq\beta$. By direct Laplace inversion we have \cite{abramowitz1965handbook}
\begin{equation}
\resizebox{0.9\hsize}{!}{$
F(t)=\begin{cases}
\begin{aligned}
&\frac{e^{-2\alpha t}-1}{2\alpha^2\sqrt{\pi t^3}}+\frac{2}{\alpha\sqrt{\pi t}}  \quad &\text{for}& \quad \alpha=\beta \\
\hfill\\
&\frac{\beta^2}{\alpha\sqrt{(\alpha-\beta)^3}}e^{-\frac{\alpha^2}{\alpha-\beta}t}\left[\operatorname{erfi}\left({\sqrt{\frac{\beta^2}{\alpha-\beta}t}}\right)-\operatorname{erfi}\left({\sqrt{\frac{\alpha^2}{\alpha-\beta}t}}\right)\right]+\frac{\alpha-\beta e^{-(\alpha+\beta)t}}{\alpha(\alpha-\beta)\sqrt{\pi t}}  \quad &\text{for}& \quad \alpha\neq\beta,
\end{aligned}
\end{cases}$}\label{generalF}
\end{equation}
where $\operatorname{erfi}(x)=-i \operatorname{erf}(ix)$ is the imaginary error function. In the main text, the exact first hitting time distribution (FHTD) are evaluated numerically using Eq. (\ref{convolution}), (\ref{generalG}) and (\ref{generalF}).



\providecommand{\noopsort}[1]{}\providecommand{\singleletter}[1]{#1}%

\end{document}